\documentclass[dvips,aos]{imsart}

\usepackage[dvips]{graphicx}
\usepackage{float,latexsym,amsmath,amsthm,amsfonts,natbib,float}
\usepackage[psamsfonts]{amssymb}
\usepackage{bbm}
\usepackage{color}

\newtheorem{theorem}{Theorem}[section]
\newtheorem{lemma}[theorem]{Lemma}
\newtheorem{proposition}[theorem]{Proposition}
\numberwithin{equation}{section}
\DeclareMathOperator{\diag}{diag}
\DeclareMathOperator{\tr}{trace}
\DeclareMathOperator{\spec}{spec}
\DeclareMathOperator*{\argmin}{argmin}
\DeclareMathOperator{\V}{V}
\newcommand{\E}{\mathbbm{E}} 
\newcommand{\R}{\mathbbm{R}}

\begin{document}

\begin{frontmatter}
\title{Recursive Bias Estimation and $L_2$ Boosting}
\runtitle{Recursive bias estimation}

\begin{aug}
\author{\fnms{Pierre-Andr\'e} \snm{Cornillon}\ead[label=e1]{pierre-andre.cornillon@supagro.inra.fr}},
\author{\fnms{Nicolas} \snm{Hengartner}\ead[label=e2]{nickh@lanl.gov}}
\and
\author{\fnms{Eric} \snm{Matzner-L\o ber}\corref{}\ead[label=e3]{eml@uhb.fr}}
\affiliation{Montpellier SupAgro, University Rennes 2 and Los Alamos National Laboratory}

\address{Address of P-A Cornillon\\
UMR ASB - Montpellier SupAgro\\
34060 Montpellier Cedex 1\\
\printead{e1}}

\address{Address of N. Hengartner\\
Los Alamos National Laboratory,\\
NW, USA\\
\printead{e2}
}

\address{Address of E. Matzner-L\o ber\\
Statistics, IRMAR UMR 6625,\\
Univ. Rennes 2, \\
35043 Rennes, France\\
\printead{e3}\\
}

\end{aug}

\begin{abstract}
This paper presents a general iterative bias correction procedure for 
regression smoothers. This bias reduction schema is shown to correspond 
operationally to the $L_2$ Boosting algorithm and provides a new statistical 
interpretation for $L_2$ Boosting. We analyze the behavior of the Boosting 
algorithm applied to common smoothers $S$ which we show depend on the 
spectrum of $I-S$. We present examples of common smoother for which 
Boosting generates a divergent sequence.  The statistical interpretation 
suggest combining algorithm with an appropriate stopping rule for the 
iterative procedure. Finally we illustrate the practical finite sample 
performances of the iterative smoother via a simulation study. 
simulations.
\end{abstract}


\end{frontmatter}

\section{Introduction}

Regression is a fundamental data analysis tool for uncovering functional 
relationships between pairs of observations $(X_i,Y_i), i=1,\ldots,n$. The 
traditional approach specifies a parametric family of regression functions to
describe the conditional expectation of the dependent variable $Y$ given the 
independent variables $X \in {\mathbb R}^p$, and estimates the free parameters 
by minimizing the squared error between the predicted values and the data.
An alternative approach is to assume that the regression function varies
smoothly in the independent variable $x$ and estimate locally the conditional 
expectation of $Y$ given $X$. This results in nonparametric regression estimators 
(e.g. \citet{fan+1996,hastie+1995,simonoff1996}).  The vector of predicted
values $\widehat Y_i$ at the observed covariates $X_i$ from a nonparametric 
regression is called a regression smoother, or simply a smoother, because the 
predicted values $\widehat Y_i$ are less variable than the original observations
$Y_i$.   

Over the past thirty years, numerous smoothers have been proposed: running-mean 
smoother, running-line smoother, bin smoother, kernel based smoother 
(\citet{nadaraya1964,watson1964}), spline regression smoother, smoothing splines 
smoother (\citet{whittaker1923,wahba1990}), locally weighted running-line 
smoother (\citet{cleveland1979}), just to mention a few.  We refer to
\citet{buja++1989,eubank1988,fan+1996,hastie+1995} for more in depth 
treatments of regression smoothers.

An important property of smoothers is that they do not require a rigid (parametric) 
specification of the regression function.  That is, we model the pairs $(X_i,Y_i)$ as

\begin{eqnarray} \label{eq:basic.model}
Y_i &=& m(X_i) + \varepsilon_i, \quad i=1,\ldots,n,
\end{eqnarray}

where $m(\cdot)$ is an unknown smooth function. The disturbances $\varepsilon_i$ 
are independent mean zero and variance $\sigma^2$ random variables that are 
independent of the covariates $X_i$, $i=1,\ldots,n$. To help our discussion on 
smoothers, we rewrite Equation (\ref{eq:basic.model}) compactly in vector form by 
setting $Y=(Y_1,\ldots,Y_n)^t$, $m=(m(X_1),\ldots,m(X_n))^t$ and 
$\varepsilon=(\varepsilon_1,\ldots,\varepsilon_n)^t$, to get 

\begin{eqnarray}
Y &=& m + \varepsilon.   \label{eq:model.vector}
\end{eqnarray}

Finally we write $\widehat m = \widehat Y =(\widehat Y_1,\ldots,\widehat Y_n)^t$, 
the vector of fitted values from the regression smoother at the observations. 
Operationally, linear smoothers can be written as

\begin{eqnarray*}
\widehat m = S_\lambda Y,
\end{eqnarray*}

where $S_\lambda$ is a $n \times n$ smoothing matrix. While in general the smoothing 
matrix will be not be a projection, it is usually a contraction (\citet{buja++1989}).  
That is,  $\|S_\lambda Y \| \leq \|Y\|$.

Smoothing matrices $S_\lambda$ typically depend on a tuning parameter, which denoted 
by $\lambda$, that governs the tradeoff between the smoothness of the estimate and the 
goodness-of-fit of the smoother to the data.  We parameterize the smoothing matrix  
such that large values of $\lambda$ will produce very smooth curves while small 
$\lambda$ will produce a more wiggly curve that wants to interpolate the data.  The 
parameter $\lambda$ is the bandwidth for kernel smoother,  the span size for 
running-mean smoother, bin smoother, and the penalty factor $\lambda$ for spline 
smoother. 

Much has been written on how to select an appropriate smoothing parameter,
see for example (\citet{simonoff1996}).  Ideally, we want to choose the smoothing 
parameter $\lambda$ to minimize the expected squared prediction error.  But without
explicit knowledge of the underlying regression function, the prediction 
error can not be computed.  Instead, one minimizes estimates
of the prediction error using Stein Unbiased Risk 
Estimate or Cross-Validation (\citet{li1985}).

This paper takes a different approach. Instead of selecting the tuning
parameter $\lambda$, we fix it to some reasonably large value, in a
way that ensures that the resulting smoothers \textit{oversmooths} the
data, that is, the resulting smoother will have a relatively small
variance but a substantial bias. Observe that the conditional
expectation of the $-R=-(Y-\widehat Y)$ given $X$ is the bias of the
smoother. This provides us with the opportunity of estimating the bias
by smoothing the residuals $R$, thereby enabling us to bias correct
the initial smoother by subtracting from it the estimated bias.  The
idea of estimating the bias from residuals to correct a pilot
estimator of a regression function goes back to the concept of
\textit{twicing} introduced by (\citet{tukey1977}) to estimate bias
from model misspecification in multivariate regression.  Obviously,
one can iteratively repeat the bias correction step until the increase
to the variance from the bias correction outweighs the magnitude of
the reduction in bias, leading to an iterative bias correction.

\medskip

Another iterative function estimation method, seemingly unrelated to bias 
reduction, is Boosting. Boosting was introduced as a machine learning
algorithm for combining multiple weak learners by averaging their
weighted predictions (\citet{schapire1990,freund1995}).  The good
performance of the Boosting algorithm on a variety of datasets
stimulated statisticians to understand it from a statistical point of
view.  In his seminal paper, \citet{breiman1998} shows how Boosting
can be interpreted as a gradient descent method. This view of Boosting
was reinforced by \citet{friedman2001}. Adaboost, a popular variant of the
Boosting algorithm, can be understood as a method for fitting an
additive model (\citet{friedman++2000}) and recently \citet{efron+++2004} 
made a connection between $L_2$ Boosting and Lasso for linear models.

But connections between iterative bias reduction and Boosting can be
made. In the context of nonparametric density estimation,
\citet{marzio+2004} have shown that one iteration of the Boosting
algorithm reduced the bias of the initial estimator in a manner
similar to the multiplicative bias reduction methods
(\citet{hjort+1995,jones++1995,hengartner+2007}).  In the follow-up
paper (\citet{marzio+2007}), they extend their results to the
nonparametric regression setting and show that one step of the
Boosting algorithm applied to an oversmooth effects a bias reduction.
As expected, the decrease in the bias comes at the cost of an increase
in the variance of the corrected smoother.

In Section 2, we show that in the context of regression, such iterative bias 
reduction schemes obtained by correcting an estimator by smoothers of the residuals 
correspond operationally to the $L_2$ Boosting algorithm.  This provides a novel
statistical interpretation of $L_2$ Boosting.  This new interpretation helps 
explain why, as the number of iteration increases, the estimator eventually 
deteriorates. Indeed, by iteratively reducing the bias, one eventually adds more 
variability than one reduces the bias.

In Section 3, we discuss the behavior
of the $L_2$ Boosting of many commonly used smoothers: smoothing splines,
Nadaraya-Watson kernel and $K$-nearest neighbor smoothers.   Unlike the good behavior
of the $L_2$ boosted smoothing splines discussed in \citet{buhlmann+2003}, 
we show that Boosting $K$-nearest neighbor
smoothers and kernel smoothers that are not positive definite produces a sequence 
of smoothers that behave erratically after a small number of iteration, and eventually
diverge.  The reason for the failure of the $L_2$ Boosting algorithm, when applied to
these smoothers, is that the bias is overestimated. As a result, the Boosting
algorithm over-corrects the bias and produces a divergent smoother sequence. 
Section 4 discusses modifications to the original smoother to ensure good 
behavior of the sequence of boosted smoothers.

To control both the over-fitting and over-correction problems,  one needs to 
stop the $L_2$ Boosting algorithm in a timely manner.  Our interpretation of the $L_2$
Boosting as an iterative bias correction scheme leads us to propose in Section 5 
several data driven stopping rules: Akaike Information Criteria (AIC), a modified AIC, 
Generalized Cross Validation (GCV), one and $L$-fold Cross Validation, and 
estimated prediction error estimation using data splitting.    Using either the
asymptotic results of \citet{li1987} or the finite sample oracle inequality of 
\citet{hengartner++2002}, we see that stopped boosted smoother has desirable
statistical properties.   We use either of these theorems to conclude that the desirable
properties of the boosted smoother does not depend on the initial pilot smoother, provided 
that the pilot oversmooths the data.  This conclusion is reaffirmed from the simulation
study we present in Section 6.
To implement these data driven stopping rules, we need to calculate 
predictions of the smoother for any desired value of the covariates, 
and not only at the observations. We show in Section 5 how to extend linear 
smoothers to give predictions at any desired point.  

The simulations 
in Section 6 show that when we combine a GCV based stopping rule to the $L_2$ 
Boosting algorithm seems to work well.  It stops early when the Boosting algorithm 
misbehaves, and otherwise takes advantage of the bias reduction.  Our simulation 
compares optimum smoothers and optimum iterative bias corrected smoothers (using
generalized cross validation) for general smoothers without knowledge
of the underlying regression function.  We conclude
that the optimal iterative bias corrected smoother outperforms the 
optimal smoother. 

Finally, the proofs are gathered in the Appendix.

\section{Recursive bias estimation} \label{section:bias}
In this section, we define a class of iteratively bias corrected linear smoothers and highlight some of their properties.

\subsection{Bias Corrected Linear Smoothers}
For ease of exposition, we shall consider the univariate 
nonparametric regression model in vector form 
(\ref{eq:model.vector}) from Section 1
\begin{eqnarray*}
Y &=& m + \varepsilon,
\end{eqnarray*}
where the errors $\varepsilon$ are independent, have mean zero and constant 
variance $\sigma^2$, and are independent of the covariates $X=(X_1,\ldots,X_n)$,
$X_j \in {\mathbb R}$.  Extensions to multivariate smoothers are strait forward
and we refer to \citet{buja++1989} for example.

Linear smoothers can be written as
\begin{equation} \label{eq:smoother.0}
\widehat m_1 = S Y,
\end{equation}
where $S$ is an $n \times n$ smoothing matrix.  Typical smoothing matrices
are contractions, so that $\|S Y\| \leq \|Y\|$, and as a result the associated smoother
$SY$ is called a shrinkage smoother (see for example \citet{buja++1989}).
Let $I$ be the $n \times n$ identity matrix.

The linear smoother 
(\ref{eq:smoother.0}) has bias
\begin{equation} \label{eq:bias.0}
B(\widehat m_1) = {\mathbb E}[\widehat m_1|X] - m = (S-I)m
\end{equation}
and variance 
\begin{eqnarray*}
V(\widehat m_1|X) = S S^\prime \sigma^2,
\end{eqnarray*}
respectively.   

A natural question is ``how can one estimate the bias?''
To answer this question, observe
that the residuals $R_1=Y-\widehat m_1=(I-S)Y$ 
have expected value 
${\mathbb E}[R_1|X] = m - {\mathbb E}[\widehat m_1|X] = (I-S)m = -B(\widehat m_1)$.
This suggests estimating the bias by smoothing the negative residuals
\begin{equation} \label{eq:bias.1}
\widehat b_1 := -SR_1 = -S(I-S)Y.
\end{equation}
This bias estimator is zero whenever the smoothing matrix 
$S$ is a projection, as is the case for linear regression, bin smoothers and
regression splines.  However, since most common smoothers
are not projections, we have an opportunity to extract further
signal from the residual and possibly improve upon the initial
estimator.

Note that a smoothing matrix other than $S$ can be used to estimate the bias in 
(\ref{eq:bias.1}),  but as we shall see, in many examples,
using $S$ works very well, and leads to an attractive interpretation
of Equation (\ref{eq:bias.1}).  Indeed, since the matrices $S$ and
$I-S$ commute, we can express the estimated bias as 
\begin{eqnarray*}
\widehat b_1 = -S(I-S)Y = -(I-S)SY = (S-I) \widehat m_1.
\end{eqnarray*}
We recognize the latter as the right-hand side of (\ref{eq:bias.0}) with 
the smoother $\widehat m_1$ replacing the unknown vector $m$.
This says that $\hat b_1$ is a plug-in estimate for the bias $B(\hat m_1)$.

Subtracting the estimated bias from the initial smoother $\widehat m_1$
produces the \textit{twicing} estimator
\begin{eqnarray*}
\widehat m_2 & = & \widehat m_1 - \widehat b_1\\
&=& (S + S(I-S))Y \\
&=& (I - (I-S)^2)Y.
\end{eqnarray*}
Since the twiced smoother $\widehat m_2$ is also a linear smoother,
one can repeat the above discussion with 
$\widehat m_2$ replacing $\widehat m_1$,  producing a  \textit{thriced}
linear smoother.  We can iterate the bias correction step to recursively  define
a family of bias corrected smoothers.  Starting with $\widehat m_1 = SY$,
construct recursively for $k=2,3,\ldots$, the sequences of residuals, estimated bias
and bias corrected smoothers
\begin{eqnarray}
R_{k-1} &=& (I-S)^{k-1} Y \nonumber \\
\widehat b_k & = & -S R_{k-1} = -(I-S)^{k-1} SY \nonumber \\
\widehat m_k &=& \widehat m_{k-1} - \widehat b_k = \widehat m_{k-1} + SR_{k-1}. \label{eq:def.mk}
\end{eqnarray}

We show in the next theorem that the iteratively bias corrected smoother 
$\hat m_k$ defined by Equation \ref{eq:def.mk} has a nice representation in terms of the 
original smoothing matrix $S$.
\begin{theorem}\label{theorem:iterative}
The $k^{th}$ iterated bias corrected linear smoother $\widehat m_k$ (\ref{eq:def.mk})
can be explicitly written
as
\begin{eqnarray}
\widehat m_k &=& S[I+(I-S)+(I-S)^2+\dots+(I-S)^{k-1}]Y \nonumber \\
&=& [I-(I-S)^k]Y=S_k Y.\label{eq:fk}
\end{eqnarray}
\end{theorem}

\begin{figure}[H]
\begin{center}
\includegraphics{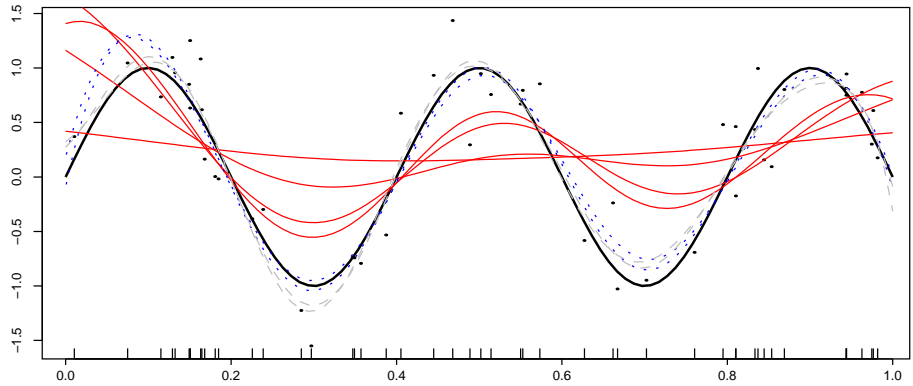}
\caption{True function $m_1$ (fat plain line) and different estimators varying with
the number of iterations $k$.\label{fig:exemple1}} 
\end{center}
\end{figure}

{\bf Example with a Gaussian kernel smoother}  
Throughout the next two sections, we shall use the following example to illustrate 
the behavior of the Boosting algorithms applied to various common smoothers.  Take 
the design points to be 50 independently drawn
points from an uniform distribution on the unit interval $[0,1]$. The true 
regression function is $m(x)=\sin(5 \pi x)$, the solid line in the Figure
\ref{fig:exemple1}, and the disturbances are mean zero Gaussians with 
variance producing a signal to noise ratio of five.

In the next figure, the initial smoother is a kernel one, with a bandwidth 
equals to 0.2 and a Gaussian kernel.  This pilot smoother heavily 
oversmooths the data, see Figure \ref{fig:exemple1} that shows that the
pilot smoother (plain line) is nearly constant.   The iterative bias corrected 
estimators are plotted 
in figure (\ref{fig:exemple1})
for values of $k$, the number of iterations, 
in $\{1,10,50,100,500,10^3,10^5,10^6\}$ 

Figure \ref{fig:exemple1} shows how each bias correction iteration changes
the smoother, starting from a nearly constant smoother and slowly
deforming (going down into the valleys and up into the peaks) with 
increasing number of iterations $k=10$, $k=50$ and $k=100$.   After
500 iterations, the iterative smoother is very close to the true function. 
However when the number of iterations is very large (here $k=10^5$ and 
$10^6$) the iterative smoother deteriorates.

\medskip

\begin{lemma}\label{MSE}
The squared bias and variance of the 
$k^{th}$ iterated bias corrected linear smoother $\widehat m_k$ (\ref{eq:def.mk})
are
\begin{eqnarray*}
B^2(\hat m_k) &=& m^t\left((I-S)^k\right)^t(I-S)^k m\\
\V(\hat m_k) &=& \sigma^2(I-(I-S)^k)\left((I-(I-S)^k)\right)^t.
\end{eqnarray*}
\end{lemma}

\bigskip
\noindent
\textbf{Remark:}  Symmetric smoothing matrices $S$
can be decomposed as  $S=P_{S}\Lambda_{S}P_{S}^t$, 
with orthonormal matrix $P_S=[u_1,u_2,\cdots,u_n]$ 
and diagonal matrix $\Lambda_S$.
\begin{eqnarray}
\label{sym.iter}
\hat m_k &=& P_{S} \diag(1-(1-\Lambda_S)^k)P_{S}^{t} Y
=  \sum_j (1-(1-\lambda_j)^k) u_ju_j^t Y.
\end{eqnarray}
Applying Lemma \ref{MSE}, we get 
\begin{eqnarray*}
B^2(\hat m_k) &=& m^tP_S(I-\Lambda_S)^{2k} P_S^tm\\
\V(\hat m_k) &=& \sigma^2P_S(I-(I-\Lambda_S)^k)^2P_S^t.
\end{eqnarray*}
Hence if the magnitude of the eigenvalues of $I-S$ are bounded 
by one,  each iteration of the bias correction will decrease the bias
and increase the variance.   This monotonicity (decreasing bias,
increasing variance) with increasing number of iterations $k$ allows us
consider data driven selection for number of bias correction
steps to achieves the best compromise between 
bias and variance of the smoother.

\medskip

The preceding remark suggests that the behavior of the iterative
bias corrected smoother $\widehat m$ is tied to the spectrum of 
$I-S$, and not of $S$.  The next theorem collects the various convergence
results for iterated bias corrected linear smoothers.

\begin{theorem} \label{theorem:converge}  Suppose that the singular values 
$\lambda_j = \lambda_j(I-S)$
of $I-S$ satisfy
\begin{eqnarray} \label{eq:condition.theorem}
-1 < \lambda_j < 1 \quad \mbox{for} \quad j=1,\ldots,n.
\end{eqnarray}
Then we have that
\begin{eqnarray*}
&& \|\hat b_k\| < \|\hat b_{k-1}\| \quad \mbox{ and } \quad \lim_{k \rightarrow \infty} 
\hat b_k = 0,\\
&& \|R_k\| < \|R_{k-1}\| \quad \mbox{ and } \quad \lim_{k \rightarrow \infty} 
R_k = 0,\\
&& \lim_{k \rightarrow \infty} \widehat m_k = Y \quad \mbox{ and } \quad 
\lim_{k \rightarrow \infty} {\mathbb E}[\|\widehat m_k - m\|^2] = n \sigma^2.
\end{eqnarray*}
Conversely, if $I-S$ has a singular value $|\lambda_j| > 1$, then
\begin{eqnarray*}
\lim_{k \rightarrow \infty} \|\hat b_k\| = \lim_{k \rightarrow \infty} \|R_k\|
= \lim_{k\rightarrow \infty} \| \widehat m_k\| = \infty.
\end{eqnarray*}
\end{theorem}

\noindent
\textbf{Remark 1:}  This theorem shows that iterating the booting
algorithm to reach the limit of the
sequence of boosted smoothers, $Y_\infty$, is not the desirable.  
However, since each iteration decreases the bias and increases the
variance, a suitably stopped Boosting estimator is likely to improve
upon the initial smoother.

\noindent
\textbf{Remark 2:}  When $|\lambda_j(I-S)| > 1$, the iterative bias 
correction fails.  The reason is that $\hat b_k$ overestimates the true bias
$b_k$,  and hence Boosting repeatedly overcorrects the bias of the 
smoothers, which results in a divergent sequence of smoothers.
Divergence of the sequence of boosted smoothers can be detected numerically, making
it possible to avoid this bad behavior by combining the iterative bias
correction procedure with a suitable stopping rule.

\noindent
\textbf{Remark 3:} The assumption that for all $j$, the singular
values $-1 < \lambda_j(I-S) < 1$ implies that $I-S$ is a contraction,
so that $\|(I-S)Y\| < \|Y\|$.  This condition does not imply that the 
smoother $S$ itself is a shrinkage smoother as defined by (\citet{buja++1989}).
Conversely, not all shrinkage estimators satisfy the condition
\ref{eq:condition.theorem} of the theorem.  In Section 3, we will 
given examples of common shrinkage smoothers 
for which $|\lambda_j(I-S)| > 1$, and show numerically that for 
these shrinkage smoothers, the iterative bias correction scheme 
will fail.

\bigskip

\subsection{$L_2$ Boosting for regression}
Boosting is one of the most successful and practical methods that arose 15 years 
ago from the machine learning community (\citet{schapire1990,freund1995}). In 
light of \citet{friedman2001}, the Boosting algorithms has been interpreted 
as functional gradient descent technique. Let us summarize the $L_2$ Boost 
algorithm described in \citet{buhlmann+2003}.

{\it Step 0:} Set $k=1$.   Given the data $\{(X_i,Y_i), i=1,\ldots,n\}$, calculate an pilot 
regression smoother 
\begin{eqnarray*}
\hat F_1(x) &=& h(x;\hat \theta_{X,Y}),
\end{eqnarray*}
by least squares fitting of the parameter, that is,
\[
\hat \theta_{X,Y}=\argmin_{\theta}\sum_{i=1}^n (Y_i-h(X_i,\theta))^2.
\]

{\it Step 1:}  With a current smoother $\widehat F_k$, 
compute the residuals $U_i=Y_i-\hat F_k(X_i)$ and fit the real-valued 
learner to the current residuals by least square. The fit is denoted by 
$\hat f_{k+1}(.)$. Update
\begin{eqnarray} \label{eq:update.gradient}
\hat F_{k+1}(.) &=& \hat F_k(.) + \hat f_{k+1}(.).
\end{eqnarray}

{\it Step 2:} Increase iteration index $k$ by one and repeat step 1.

\begin{lemma}[Buhlmann and Yu, 2003] The smoothing matrix 
associated with the $k^{th}$ Boosting iterate of linear 
smoother with smoothing matrix $S$ is
\begin{eqnarray*}
\hat F_{k} = (I-(I-S)^k)Y=B_k Y.
\end{eqnarray*}
\end{lemma}

\medskip

Viewing Boosting as a greedy gradient descent method, the update formula 
(\ref{eq:update.gradient}) is often modified to include \textit{convergence
factor} $\mu_k$,  as in \citet{friedman2001}, to become
\begin{eqnarray*}
\hat F_{k+1}(.) &=& \hat F_k(.) + \hat \mu_{k+1} \hat f_{k+1}(\cdot),
\end{eqnarray*}
where $\hat \mu_{k+1}$ is the best step toward the best direction $\hat f_{k+1}(\cdot)$.

This general formulation allows a great deal of flexibility, both in selecting the
type of smoother used in each iteration of the Boosting algorithm, and in the 
selection of the convergence factor.  For example, we may start with a running 
mean pilot smoother, and use a smoothing spline to estimate the bias in the
first Boosting iteration and a nearest neighbor smoother to estimate the bias
in the second iteration.  However in practice, one typically uses the same smoother
for all iterations and fix the convergence factor $\mu_k \equiv \mu \in (0,1)$.
That is, the sequence of smoothers resulting from the Boosting algorithm is given by
\begin{eqnarray} \label{eq:mod.boosting}
\hat F_{k} = (I-(I- \mu S)^k)Y=B_k Y.
\end{eqnarray}

We shall discuss in detail in Section \ref{ingeniering} the impact of this 
convergence factor and other modifications to the Boosting algorithm
to ensure good behavior of the sequence of boosted smoothers.

\section{Boosting classical smoothers}

This section is devoted to understanding the behavior of  
the iterative Boosting schema using classical smoothers, which  
in light of  Theorem \ref{theorem:converge},  
depends on the magnitude of the singular values of the matrix $I-S$.  

\medskip

We start our discussion by noting that
Boosting a {\bf projection type smoothers} is of no interest 
because residuals $(I-S)Y$ are orthogonal to smoother
$SY$.  It follows that the smoothed residuals $S(I-S)Y=0$, and
as a result,  $\widehat m_k = \widehat m_1$ 
for all $k$.  Hence Boosting a bin smoother or a regression spline smoother 
leaves the initial smoother unchanged.

\medskip 

Consider the {\bf $K$-nearest neighbor smoother}.
Its associated smoothing matrix is $S_{ij}=1/K$ if $X_j$ belongs to the $K$-nearest
neighbor of $X_i$ and $S_{ij}=0$ otherwise.  Note that this smoothing
matrix is not symmetric.  While this smoother enjoys many
desirable properties, it is not well suited for Boosting because the matrix
$I-S$ has singular values larger than one.

\begin{theorem} \label{knn}
In the fixed design or in the uniform design, as soon as the number of $K$ is bigger 
than one and smaller than $n$, at least one singular value of $I-S$ is bigger than 1.
\end{theorem}

The proof of the theorem is found in the appendix.   A consequence
of Proposition \ref{knn} and Theorem \ref{theorem:converge},  is that the Boosting 
algorithm applied to a $K$-nearest neighbor smoother produces a sequence 
of divergent smoothers, and hence should not be used in practice. 

\medskip
{\bf Example continued with $K$-nearest neighbor smoother.}   
We confirm this behavior numerically.  Using the same data as before, we apply the 
Boosting algorithm starting with an pilot $K$-nearest neighbor smoother with $K=10$. 
The pilot estimator is plotted in a plain line, and the various boosted smoothers  
with $k$, the number of iterations, valued in  $\{2,\cdots,5\}$ in dotted lines.

\begin{figure}[H]
\begin{center}
\includegraphics{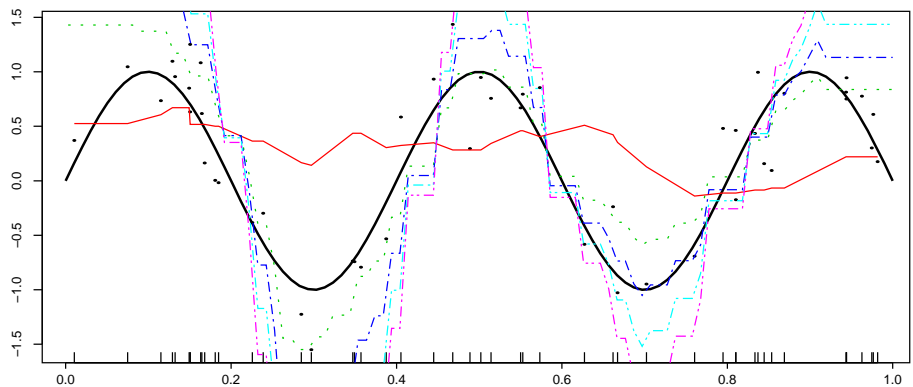}
\caption{True function $m_1$ (fat plain line) and different estimators varying with
the number of iterations $k$.\label{fig:uniform}} 
\end{center}
\end{figure}
For $k=1$, the pilot smoother is nearly constant (since we take $K=10$ neighbors)
and very quickly the iterative bias corrected estimator explodes.  Qualitatively,
the smoothers are getting higher at the peaks and lower in the valleys,
which is consistent with an overcorrection of the bias.  Contrast 
this behavior with the one shown in Figure 1.

\medskip

{\bf Kernel type smoother}. For Nadaraya kernel type estimator, the smoothing matrix 
$S$ has entries $S_{ij} =  K_h(X_i-X_j)/\sum_k K_h(X_i-X_k)$, where $K(.)$ is a symmetric 
function (e.g., uniform, Epanechnikov, Gaussian), $h$ denotes the 
bandwidth and $K_h(\cdot)$ is the scaled kernel $K_h(t)=h^{-1}K(t/h)$. The matrix $S$ is not symmetric but can be written as $S= D \mathbb{K}$ where $\mathbb{K}$ 
is symmetric with general element $[K_h(X_i-X_j)]$ and $D$ is diagonal with element 
$1/\sum_j K_h(X_i-X_j)$. Algebraic manipulations allows us to rewrite the iterated 
estimator as
\begin{eqnarray*}
\hat m_k &=& [I-(I-S)^k]Y \\
&=& [I - (D^{1/2}D^{-1/2} - D^{1/2}D^{1/2}\mathbb{K}D^{1/2}D^{-1/2})^k]Y\\
&=&[I - D^{1/2}(I - D^{1/2}\mathbb{K}D^{1/2})^kD^{-1/2}]Y\\
&=&D^{1/2}[I-(I - A)^k] D^{-1/2}Y.
\end{eqnarray*}
Since the matrix $A=D^{1/2}\mathbb{K}D^{1/2}$ is symmetric, we apply the classical decomposition
$A=P_{A}\Lambda_AP_A^{t}$, with $P_A$ orthonormal and $\Lambda_A$ diagonal, to 
get a closed form expression for the boosted smoother 
\begin{eqnarray*}
\hat m_k &=&D^{1/2}P_{A}[I-(I - \Lambda_A)^k]P_A^{t}D^{-1/2}Y.
\end{eqnarray*}
The eigen decomposition of $A=D^{1/2}\mathbb{K}D^{1/2}$ can be used to describe the 
behavior of the sequence of iterative estimators. In particular, any eigenvalue 
of $A=D^{1/2}\mathbb{K}D^{1/2}$  that is negative or greater than 2 will lead to 
unstable procedure. If the kernel $K(\cdot)$ is a symmetric probability density 
function positive definite, then the spectrum of the Nadaraya-Watson kernel 
smoother lies between zero and one.

\begin{theorem} \label{kernel}
If the inverse Fourier-Stieltjes transform of a kernel $K(\cdot)$ 
is  a real positive finite measure,  then the 
spectrum of the Nadaraya-Watson kernel smoother 
lies between zero and one.

Conversely, suppose that $X_1,\ldots,X_n$ are an independent 
$n$-sample from a density $f$ (with respect to Lebesgue measure)
that is bounded away from zero on 
a compact set strictly included in the support of $f$.
If the inverse Fourier-Stieltjes transform of a 
kernel $K(\cdot)$ is not a positive finite measure, then with probability
approaching one as the sample size $n$ grows to infinity, the maximum
of the spectrum of $I-S$ is larger than one.
\end{theorem}

\noindent
{\bf Remark 1:} Since the $\spec(A)$ is the same as the $\spec(S)$ and
$S$ is row stochastic, we conclude that $\spec(A) \leq 1$. So we are
only concern by the presence of negative eigenvalues in the spectrum
of $A$.  

\noindent
{\bf Remark 2:} 
In \citet{marzio+2008} proved the first part of the theorem.  Our
proof of the converse shows that for large enough sample sizes,
most configurations from a random design lead to smoothing matrix
$S$ with negative singular values.   

\noindent
{\bf Remark 3:} The assumption that the inverse Fourier-Stieltjes 
transform of a kernel $K(\cdot)$  is  a real positive finite measure
is equivalent to the kernel $K(\cdot)$ being positive a definite function, 
that is, for any finite set of points $x_1,\ldots,x_m$, the matrix
\[
\left ( \begin{array}{ccccc}
K(0) & K(x_1-x_2) & K(x_1-x_3) & \dots & K(x_1-x_m) \\
K(x_2-x_1) & K(0) & K(x_2-x_3) & \dots & K(x_2-x_m)\\
\vdots & & & & \vdots \\
K(x_m-x_1) & K(x_m - x_2) & K(x_m-x_3) & \dots & K(0)
\end{array}
\right )
\]
is positive definite.   We refer to \citet{schwartz1993} for a detailed
study of positive definite functions.

\medskip

The Gaussian and triangular kernels are positive definite kernels 
(they are the Fourier transform of a finite positive measure 
(\citet{feller1966})) and in light of Theorem \ref{kernel} the 
Boosting of Nadaraya-Watson kernel smoothers with these kernels 
produces a sequence of well behavior smoother. However, the uniform and 
the Epanechnikov kernels are not positive definite.  Theorem \ref{kernel}
states that for large samples, the spectrum of $I-S$ is larger than 
one and as a result the sequence of boosted smoother diverges. 
Proposition \ref{unif} below strengthen this result by 
stating that the largest singular value of $I-S$ is always larger 
than one.

\begin{proposition}
\label{unif}
Let $S$ be the smoothing matrix of a 
Nadaraya-Watson regression smoother based on either the uniform or the
Epanechnikov kernel.  Then the largest singular value of $I-S$ is 
larger than one.
\end{proposition}

\medskip
{\bf Example continued with Epanechnikov kernel smoother.} 
In the next figure, the pilot smoother is a kernel one with an Epanechnikov 
kernel and with bandwidth is equal to 0.15.  The pilot smoother is the plain 
line, and the subsequent iterations with $k$, the number of  iterations, 
valued in $\{1,2,5,10,20,50,100\}$, are the  dotted lines. 

\begin{figure}[H]
\begin{center}
\includegraphics{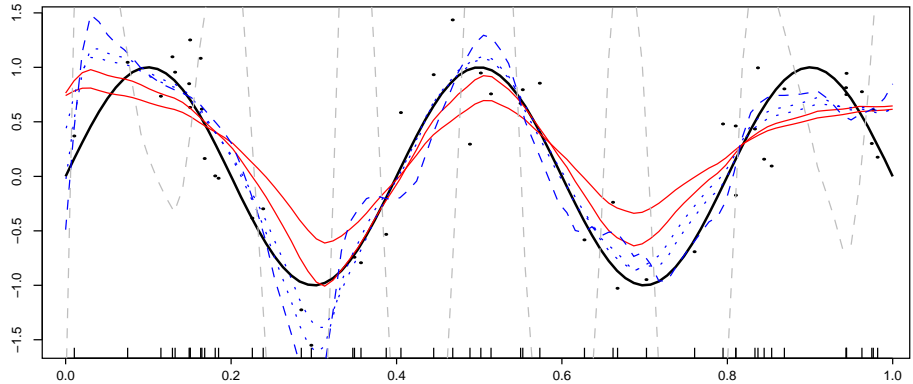}
\caption{True function $m_1$ (fat plain line) and different estimators varying with
the number of iterations $k$.\label{fig:knn}} 
\end{center}
\end{figure}
For $k=1$, the pilot smoother oversmooths the true regression since the bandwidth 
takes almost  one third of the data and very quickly the iterative estimator explodes.
Contrast this behavior with the one shown by the Gaussian kernel smoother in 
Figure \ref{fig:exemple1}.

\medskip

Finally, let us now consider the {\bf smoothing splines smoother}. The smoothing 
matrix $S$ is  symmetric,  and therefore admits an eigen decomposition.
Denote by $\{u_1,u_2,\cdots,u_n\}$ an orthonormal basis of eigenvectors of 
$S$ associated to the eigenvalues $1 \geq \lambda_1 \geq \lambda_2 \geq \cdots \geq \lambda_n
\geq 0$ (\citet{utreras1983}). Denote by $P_{S}=[u_1,u_2,\cdots,u_n]$ the 
orthonormal matrix of column eigenvectors and write 
$S= P_{S} \diag(\lambda_S)P_{S}^{t}$, that is $S= \sum_j \lambda_j u_ju_j^t$.
The iterated bias reduction estimator is given by (\ref{sym.iter}). 
Since all the eigenvalues are between 0 and 1, then if $k$ is large, 
the iterative procedure kills the eigenvalues less than 1 and put the others to 1.

\medskip

\begin{figure}[H]
\begin{center}
\includegraphics{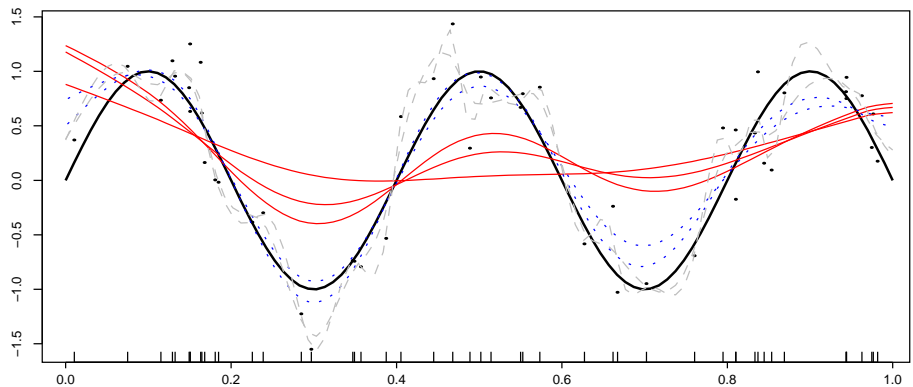}
\caption{True function $m_1$ (fat plain line) and different estimators varying with
the number of iterations $k$.\label{fig:exemple2}} 
\end{center}
\end{figure}

{\bf Example continued with smoothing splines smoother}  
In the next figure, the pilot smoother is a smoothing spline, with $\lambda$
equals to 0.2. The different estimators are plotted 
in figure (\ref{fig:exemple2}), with the pilot estimator in plain line and 
the boosted  smoothers with number of iterations $k$  being $\{10, 50, 100,
500, 10^3, 10^5,10^6\}$
in dotted lines.
The pilot estimator is more variable than the pilot estimator of figure 
\ref{fig:exemple1} and by the way the convergence and the deterioration 
arise faster.
\section{Smoother engineering}\label{ingeniering}

Practical implementations of the Boosting algorithm include a user selected
convergence factor $\mu \in (0,1)$  that appears in the definition of the boosted 
smoother 
\begin{eqnarray} \label{eq:def.Bk}
\hat m_{k} = (I-(I- \mu S)^k)Y=B_k Y.
\end{eqnarray}
In this section, we show that when $\mu < 1$, one effectively operates a 
partial bias correction.  This partial bias correction does not however resolve the problems
associated with Boosting a nearest neighbor or Nadaraya Watson kernel smoothers with
compact kernel we exhibited in the previous section.    To resolve these problems,
we propose to suitably modify the boosted smoother.  We call such targeted changes 
\textit{smoother engineering}.

\medskip

The following iterative partial bias reduction scheme is equivalent to the
Boosting algorithm defined by Equation (\ref{eq:def.Bk}):
Given a smoother $\widehat m_k = B_k Y$ at the $k^{th}$ iteration 
of the Boosting algorithm, 
calculate the residuals $R_k$ and estimated bias $\widehat b_k$
\begin{eqnarray*}
R_k &=& Y - \widehat m_k = (I-B_k)Y\\
\widehat b_k &=& S R_k = S(I-B_k) Y.
\end{eqnarray*} 
Next, given $0 < \mu < 1$, consider the partially bias corrected smoother
\begin{equation} \label{eq:m.k+1}
\widehat m_{k+1} = \widehat m_k + \mu \widehat b_k.
\end{equation}
Algebraic manipulations of the smoothing matrix of the right-hand side of 
(\ref{eq:m.k+1}) yields
\begin{eqnarray*}
B_k + \mu S(I-B_k) = I - (I-\mu S)^{k+1},
\end{eqnarray*}
from which we conclude that $\widehat m_{k+1}$ satisfies (\ref{eq:def.Bk}) 
and therefore is the $(k+1)^{th}$ iteration of the Boosting algorithm. 
It is interesting to rewrite (\ref{eq:m.k+1}) as
\begin{eqnarray*}
\widehat m_{k+1} = (1-\mu) \widehat m_k + \mu \left [ \widehat  m_k + \widehat b_k \right ],
\end{eqnarray*}
which shows that boosted smoother $\widehat m_{k+1}$ is a convex combination 
between the smoother $\widehat m_k$ at iteration $k$, and the fully bias 
corrected smoother $\widehat m_k+\widehat b_k$. As a result, we understand how 
the introduction of a convergence factor produces  a "weaker learner" than 
the one obtained for $\mu=1$.

In analogy to Theorem \ref{theorem:converge}, the behavior of the sequence of 
the smoother depends on the spectrum of $I-\mu S$.  Specifically, if 
$\max_j |\lambda_j(I-\mu S)| \leq 1$, then $\lim_{k \longrightarrow \infty} \widehat m_k = Y$,and conversely, if $\max_j |\lambda_j(I-\mu S)| > 1$, 
$\lim_{k \longrightarrow \infty} \|\widehat m_k\| = \infty$.  Inspection of the 
proofs of propositions \ref{knn} and \ref{kernel} reveal that the spectrum 
of $(I-\mu S)$ for both the nearest neighbor smoother and the Nadaraya Watson 
kernel smoother has singular values of magnitude larger than one. Hence the 
introduction of the convergence factor does not help resolve the difficulties 
arising when Boosting these smoothers.

To resolve the potential convergence issues, one needs to suitably modify the 
underlying smoother to ensure that the magnitude of the singular values of 
$I-\mu S$ are bounded by one. A practical solution is to replace the smoothing 
matrix $S$ by $S^\star = S S^t$.  If $S$ is a contraction, it follows that the 
eigenvalues of $I-S^\star$ are nonnegative and bounded by one.  Hence the 
Boosting algorithm with this smoother will produce a well behaved sequence of
smoothers with $\lim_{k \longrightarrow \infty} \widehat m_k = Y$.

While substituting the smoother $S^\star$ for $S$ can produces better boosted 
smoothers in cases where Boosting failed,  our numerical experimentations has 
shown that the performance of Boosting $S^\star$ is not as good as Boosting 
$S$ when the pilot estimator enjoyed good properties, as is the case for 
smoothing splines and the Nadaraya Watson kernel smoother with Gaussian kernel.

\section{Stopping rules\label{stop}}

Theorem \ref{theorem:converge} in Section \ref{section:bias} states that the 
limit of the sequence of boosted smoothers is either the  raw data $Y$ or has 
norm $\|Y_\infty\|=\infty$. It follows that iterating the Boosting algorithm 
until convergence is not desirable. However, since each iteration of the 
Boosting algorithm reduces the bias and increases the variance, often a few 
iteration of the Boosting algorithm will produce a better smoother than the 
pilot smoother. This brings up the important question of how to decide when
to stop the iterative bias correction process.   

Viewing the latter question as a model selection problem suggests  stopping rules based
on Mallows' $C_p$ (\citet{mallows1973}),   Akaike Information Criterion, AIC,
(\citet{akaike1973}), Bayesian Information Criterion, BIC, 
(\citet{schwarz1978}), and Generalized cross validation (\citet{craven+1979}).  
Each of these selectors estimate the optimum number of iterations $k$ 
of the Boosting algorithm by minimizing estimates of the expected squared prediction 
error of the smoothers over some pre-specified set ${\mathcal K}=\{1,2,\ldots,M\}$.

Three of the six criteria we study numerically in Section 6 use plug-in 
estimates for the squared bias and variance of the expected prediction 
mean square error.   Specifically,  consider
\begin{eqnarray}
\hat k_{AIC}&=&\argmin_{k \in \mathcal{K}}
\left\{ \widehat{\sigma}^2+ 2\frac{\tr(S_k)}{n} \right\},\label{AIC}\\
&& \nonumber\\
\hat k_{GCV}&=&\argmin_{k \in \mathcal{K}}\left\{ \log{\widehat
{\sigma}^2}-2\log{\left(1-\frac{\tr(S_k)}{n}\right)} \right\},\label{GCV}\\
&& \nonumber\\
\hat k_{AIC_C} &=& \argmin_{k \in \mathcal{K}}
\left\{\log{\widehat {\sigma}^2}+1+\frac{2(\tr(S_k)+1)}
{n-\tr(S_k)-2}\right\}.\label{AICc}
\end{eqnarray}
In nonparametric smoothing, the AIC criteria (\ref{AIC}) has a noticeable 
tendency to select more iterations than needed, leading to a final smoother 
$\widehat m_{\widehat k_{AIC}}$ that typically undersmooths the data.  As a 
remedy, \citet{hurvich++1998} introduced a corrected version of the AIC
(\ref{AICc})  under the simplifying assumption that the nonparametric 
smoother $\hat m$ is unbiased, which is rarely hold in practice and which 
is particularly not true in our context. 

The other three criteria considered in our simulation study in Section 6 are 
Cross-Validation, L-fold cross-validation and data splitting, all of which 
estimate empirically the expected prediction mean square error by splitting 
the data into learning and testing sets. Implementation of these criterion 
require one to evaluate the smoother at locations outside the of the design. 
To this end, write the $k^{th}$ iterated smoother as a $k$ times bias corrected 
smoother
\begin{eqnarray*}
\widehat m_k &=& \widehat m_0 + \widehat b_1 + \dots + \widehat b_k \\
&=& S[I+(I-S)+(I-S)^2+\dots+(I-S)^{k-1}]Y,
\end{eqnarray*}
which we rewrite as
\begin{eqnarray*}
\widehat m_k &=& S \hat \beta_k,
\end{eqnarray*}
where 
\begin{eqnarray*}
\hat \beta_k &=& [I+(I-S)+(I-S)^2+\dots+(I-S)^{k-1}]Y \\
&=& Y + R_1 + R_2 + \dots + R_k
\end{eqnarray*}
is  a vector of size $n$.   Given the vector $S(x)$  of size $n$  whose
entries are the weights for predicting $m(x)$, we calculate
\begin{eqnarray}
\label{entoutx}
\widehat m_k(x) = S(x)^t \widehat \beta_k.
\end{eqnarray}
This formulation is computationally advantageous because the vector of weights 
$S(x)$ only needs to be computed once, while each Boosting iteration updates
the parameter vector $\widehat \beta_k$ by adding the residuals 
$R_k = Y-\widehat m_k$ of the fit to the previous value of the parameter, 
i.e., $\widehat \beta_k = \widehat \beta_{k-1} + R_k$. 
The vector $S(x)$ is readily computed for many of the smoothers used in practice.
For kernel smoothers, the $i^{th}$ entry in the vector $S(x)$ is
\begin{eqnarray*}
S_i(x) = \frac{K_h(x-X_i)}{\sum_j K_h(x-X_j)}.
\end{eqnarray*}
For smoothing spline, let $N(x)$ denote the vector of basis function evaluated at
$x$.  One can show that $\hat m_k(x) = N(x) M \hat \beta_k$, where $M$ 
is the $n \times n$ matrix given by
\begin{eqnarray*}
M &=& (N^tN + \lambda \Omega)^{-1} N^t.
\end{eqnarray*}
Finally, for the $K$-nn smoother, the entries of the vector $S(x)$ are
\begin{eqnarray*}
S_i(x) =  \left \{ \begin{array}{ll} 1/K & \mbox{if $X_i$ is a $K$-nn of $x$}\\
0 & \mbox{otherwise} \end{array} \right . .
\end{eqnarray*}

\bigskip

We note that if the spectrum of $I-S$ is bounded in absolute value by one,
then the parameter $\hat \beta_k \longrightarrow \beta_\infty$, and hence
we have pointwise convergence of $\widehat m_k(x)$ to some $m_\infty(x)$,
whose properties depend on $S(x)$.  

\medskip

To define the data splitting and cross validation stopping rules, one divides the
sample into two disjoint subset: a learning set ${\mathcal L}$ which is used to 
estimate the smoother $\widehat m^{\mathcal L}$, and a testing set ${\mathcal T}$ 
on which predictions from the smoother are compared to the observations.  The 
data splitting selector for the number of iterations is
\begin{eqnarray}
\label{DS}
\hat k_{DS} &=& \argmin_{k \in \mathcal{K}} \sum_{i \in \mathcal{T}} 
\left(Y_i - \hat m^{\mathcal{L}}_k(X_i) \right)^2.
\end{eqnarray}
One-fold cross validation, or simply cross validation, and more generally 
$L$-fold cross validation average the prediction error
over all partitions of the data into into learning and testing sets, with fixed 
size of the testing set $|{\mathcal T}|=L$.  This leads to 
\begin{eqnarray}
\label{CV}
\hat k_{CV} &=& \argmin_{k \in \mathcal{K}}  \sum_{|{\mathcal T}|=L} 
\sum_{i \in {\mathcal T}}^n \left(Y_i - \hat m^{\mathcal L}_k(X_i) \right)^2.
\end{eqnarray}

\medskip

We rely on the expansive literature on model selection to provide insight 
into the statistical properties of stopped boosted smoother.  For example,
Theorem 3.2 of \citet{li1987} describes the asymptotic behavior of the generalized 
cross-validation (GCV) stopping rule applied to spline smoothers.  

\begin{theorem}[Li, 1987]
Assume that Li's assumptions are verified for the smoother $S$.
Then 
\begin{eqnarray*}
\frac{\| m- S_{\hat k_{GCV}} Y \|^2}
{\inf_{k \in \mathcal{K}}\| m- S_k Y\|^2} 
\rightarrow 1 \quad \hbox{in probability}.
\end{eqnarray*}
\end{theorem}

Results on the finite sample performance for data splitting for arbitrary smoothers
is presented in Theorem 1 of \citet{hengartner++2002} who proved the following 
oracle inequality.

\begin{theorem}
For each $k$ in $\mathcal{K}$, $\lambda>0$ and $\alpha>0$, we have
\begin{eqnarray*}
P\left\{\frac{1}{m} \sum_{i=n+1}^{n+m} (\hat{m}_{K_{DS}}-m)^2(X_i)
-(1-\alpha)\sum_{i=n+1}^{n+m} (\hat{m}_{k}-m)^2(X_i) \geq \lambda \right\}\\
\leq |K| \sqrt{
\left(\frac{32(1+\alpha)\sigma^2}{\pi \alpha m \lambda}\right)
\left[\exp\left(\frac{\alpha m \lambda}{8(1+\alpha)\sigma^2}\right)-1\right]^{-1}}.
\end{eqnarray*}
\end{theorem}

\medskip

{\bf Example continued with smoothing splines}

\begin{figure}[H]
\begin{center}
\includegraphics{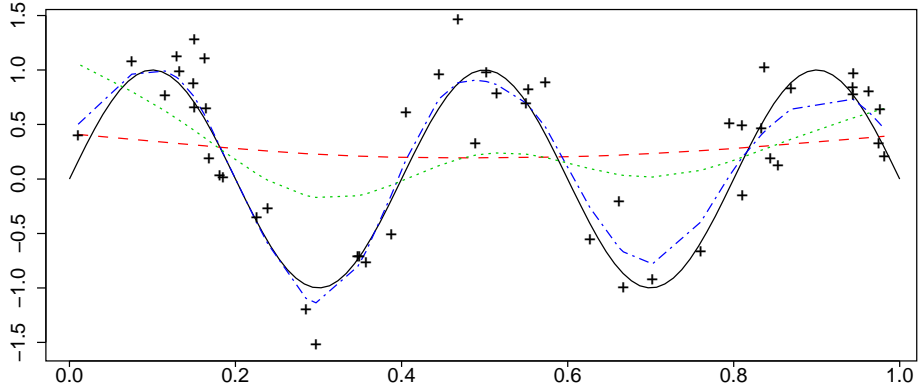}
\caption{True function $m_1$ (plain line) and different pilot smoothing spline
  smoother, $S(\lambda_1)$ (dotted line), $S(\lambda_2)$ (dashed
  line),$S(\lambda_3)$ (dash-dotted line) for the 50 data points of
  one replication (Gaussian error).\label{fig:initialsmoother}}
\end{center}
\end{figure}

Figure \ref{fig:initialsmoother} shows the
three pilot smoothers (smoothing splines with different smoothing parameters)
considered in the simulation study in Section 6.

Starting with the smoothest pilot smoother $S(\lambda_1)$, the Generalized Cross 
Validation criteria stops after 1389 iterations.  Starting with smoother 
$S(\lambda_2)$, GCV stopped after 23 iterations, while starting with the noisiest 
pilot $S(\lambda_3)$, GCV stopped after one iteration.
It is remarkable how similar the final smoother are.  

\begin{figure}[H]
\begin{center}
\includegraphics{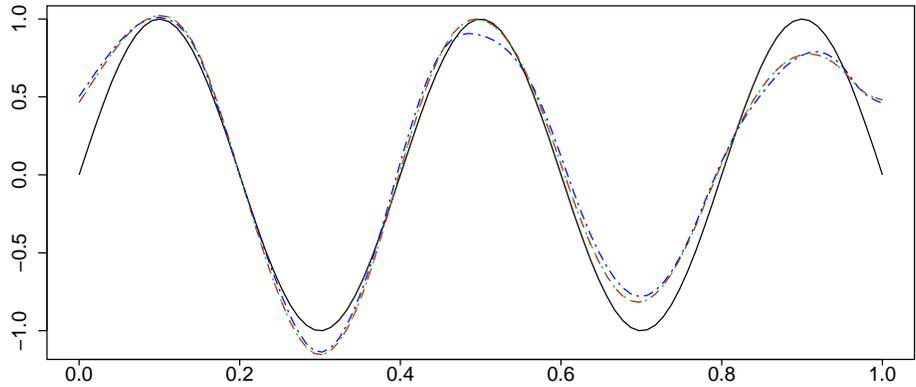}
\caption{True function $m_1$ (plain line) and different pilot smoothing spline
  smoother, $S(\lambda_1)$ (dashed
  line), $S(\lambda_2)$ (dotted line),$S(\lambda_3)$ (dash-dotted line) for the same 50 data points as in figure \ref{fig:initialsmoother} of
  one replication (Gaussian error).\label{fig:finalsmoother}} 
\end{center}
\end{figure}

The final selected estimators are very close to one another, despite the
different pilot smoothers and the different numbers iterations that were
selected by the GVC criteria.
Despite the flatness of the pilot smoother $S(\lambda_1)$, it
succeeds after 1389 iteration to capture the signal. 
Note that larger smoothing parameter $\lambda$ are associated 
to weaker learners that require a larger number of bias correction 
iterations before they become desirable smoothers according the
the generalized cross validation criteria.
A close examination of figure \ref{fig:finalsmoother} shows that using
the less biased estimator $S(\lambda_3)$ leads to the worse final
estimator. This can be explained as follows: if the pilot smoother
is not enough biased, after the first step almost no signal is left in
the residuals and the iterative bias reduction is stopped. 

We remark again that one does not need to keep the same
smoother throughout the iterative bias correcting scheme.  We conjecture
that there are advantages to using weaker smoothers later in the
iterative scheme, and shall investigate this in a forthcoming paper.

\section{Simulations}
This section presents selected results from our simulation study that investigates 
the statistical properties of the various data driven stopping rules. The 
simulations examine, within the framework set by \citet{hurvich++1998},  the 
impact on performance of various stopping rules, smoother type, smoothness of 
the pilot smoother, sample size, true regression function, and the relative
variance of the errors as measured by the signal to noise ratio.

We examine the influence of 
various factors on the performance of the selectors, with 100 simulation 
replications and a random uniform grid in $[0,1]$. The  error standard 
deviation is $\sigma=0.2R_g$, where $R_g$ is the range of $g(x)$ over $x \in [0,1]$.
For each setting of factors, we have
\begin{itemize}
\item[(A)] sample size: $n=$ 50, 100 and 500
\item[(B)] the following 3 regression functions, most of which were used in earlier 
studies 
\begin{enumerate}
\item $m(x) =  \sin(5\pi x)$,
\item $m(x) = 1-48x+218x^2-315x^3+145x^4$,
\item $m(x) = \exp{(x-\frac{1}{3})}\{x<\frac{1}{3}\}+\exp[-2(x-\frac{1}{3})]\{x\geq\frac{1}{3}\}$.
\end{enumerate}
\item[(C)] error distribution: Gaussian and Student(5)
\item[(D)] pilot smoothers: smoothing splines, Gaussian kernel, $K$-nearest neighbor 
type smoother
\item[(E)] three starting smoothers: $S_1$, $S_2$ and $S_3$ by decreasing order
of smoothing. 
\end{itemize}
For each setting, we compute the ideal numbers of iterations by computing at data points $\{X_i\}_{i=1}^n$
\begin{eqnarray*}
 k_{\mathrm{opt}} &=& \argmin_{k \in \mathcal{K}} \sum_{i=1}^n \|m(X_i) - \hat m_k(X_i)\|^2.
\end{eqnarray*}
Since the results are numerous we report here a summary, focusing on
the main objectives of the paper. 

\begin{figure}[H]
\begin{center}
\includegraphics{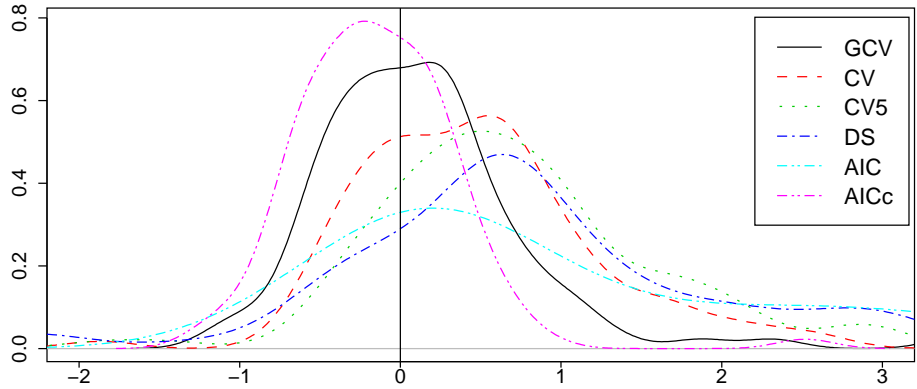}
\caption{Estimated density of $\log(\hat k/k_{\mathrm{opt}})$, $\hat
  k$ evaluated by different stopping criterion : GCV, CV (leave one
  out), CV 5 fold (CV5), data splitting (DS), AIC and corrected AIC
  (AICc). Density is estimated on 100 replication for function $m_1$,
  with Gaussian error, spline smoother $S_2$ and
  $n=50$ data points.\label{fig:densityk:kopt}}
\end{center}
\end{figure}

\medskip 
First of all, does the stopping procedures proposed in section \ref{stop} work ? 
Figure \ref{fig:densityk:kopt} represents the kernel density estimates of the
log ratios of the number of iterations to the ideal number of iterations for 
the smoothing spline type smoother. 

Obviously, negative values indicate undersmoothing ($\hat k$ smaller than
$k_{\mathrm{opt}}$, that is not enough bias reduction) while positive
values indicate oversmoothing.   The results remain essentially unchanged
over the range of starting values, regression function and smoothers types
we considered in our simulation study.

For small data sets ($n=50$), the stopping rule based on data splitting 
produced values for $\widehat k$ that were very variable.  A similar 
observation about the variability of bandwidth selection from data splitting
was made in \citep[see][]{hengartner++2002}.    We also found that the
five fold cross validation stopping rule produced highly variable values
for $\widehat k$.
 
The AIC stopping rule 
selects values $\hat k$ that are often too big (oversmoothing) and sometimes selects 
the largest possible value of $\widehat k \in K$.  
In that cases, the curve $k$ versus $AIC$ (not
shown) indicates two minimum, a local one which is around the true
value and the global one at the boundary. This can be attributed to the fact 
that the penalty term used by AIC is too small.  The AICc criteria uses 
a larger penalty term, which leads to smaller values for $\hat k$.
In fact, the selected values are typically smaller than the optimal one. 
The penalty associated with GCV lies in 
between the AICc penalty and AIC penalty, and produces in practice
values of $\widehat k$ that are closer to the optimum than either
AIC or AICc.   Finally, the leave one out cross-validation selection rule
produces $\hat k$ that are typically larger than the optimal one.

\begin{figure}[H]
\begin{center}
\includegraphics{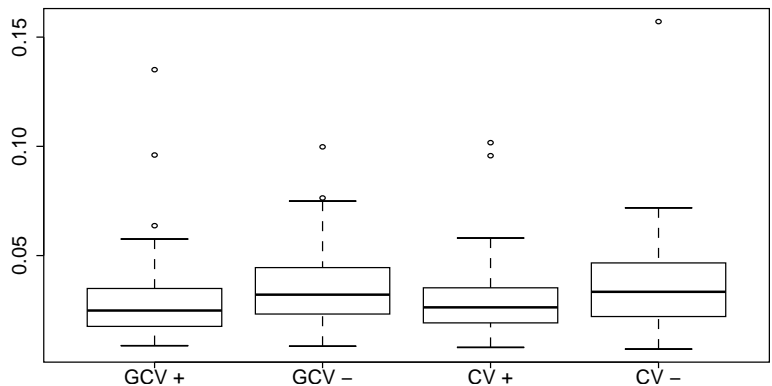}
\caption{Boxplot of $\mathrm{MSE}_{\hat m_{\hat k}}$
 when $\hat k_{\mathrm{GCV}} > k_{\mathrm{opt}}$ 
  (denoted as GCV+), of mean squared error of $\hat
  f_{\hat k_{\mathrm{GCV}}}$ when $\hat k_{\mathrm{GCV}}\le
  k_{\mathrm{opt}}$ (denoted as GCV-), and the same boxplots with
  leave one out stopping criterion. Mean squared error are estimated on a grid
of 100 points regularly spaced between 0 and 1, 100 replication for function $m_1$,
  with Gaussian error, spline smoother $S_2$ and
  $n=50$ data points.\label{fig:boxplot_stopbefore}}
\end{center}
\end{figure}

Investigation of the MSE as a function of the number of iterations $k$
reveal that, in the examples we considered, that function decreases rapidly
towards its minimum and then remains relatively flat
over a range of values to right of the minimum.  It follows that the loss
of stopping after $k_{\mathrm{opt}}$ is less than stopping before 
$k_{\mathrm{opt}}$.   We verify this empirically as follows:
for each estimate, we calculate the approximation to the integrated
mean squared error between the estimator $\hat m_{\hat k}$ and the 
true regression function $m$
\begin{eqnarray*}
\mathrm{MSE}(\hat m_{\hat k})&=& 1/100\sum_{x\in \mathcal{G}} |m(x) - \hat m_{\hat k}(x)|^2,
\end{eqnarray*}
where $\mathcal{G}$ is a fix grid of 100 regularly spaced 
points in the unit interval $[0,1]$.  We partition the calculated 
integrated mean squared error depending on whether 
$\hat k$ is bigger than $k_{\mathrm{opt}}$ or smaller than
$k_{\mathrm{opt}}$.   Figure \ref{fig:boxplot_stopbefore} presents the
boxplot of the integrated mean squared error when $\hat k$
over-estimates $k_{\mathrm{opt}}$ and when it under-estimates
$k_{\mathrm{opt}}$ and clearly shows that over-estimating
$k_{\mathrm{opt}}$ leads to smaller integrated mean squared 
error than under-estimating $k_{\mathrm{opt}}$.

For bigger data sets, say $n=100$ or bigger, most of the stopping criterion 
act the same except for the modified AIC which tends to select a smaller
number of iterations $k$ than
the ideal one. One fold cross-validation is rather computational
intensive as the usual relation between cross validated estimator at
$X_i$ and full data estimator is no longer valid \citep[e.g.][p. 47]{hastie+1995}. 

These conclusions remain true for kernel smoother and nearest neighbor
smoothers. However if the pilot smoother is not smooth enough (not biased 
enough), then the number of iteration is too small to allow us to discriminate 
between the different stopping rules.  These initial smoothers 
we name as wiggly learner are almost unbiased and therefore, 
there is little value to apply an iterative bias correction scheme. 
In conclusion, for small data sets, our simulations show that both
GCV and leave one cross-validation work well, 
and for bigger data sets, we recommend using GCV.

\medskip

Tables (\ref{tab:mse}) and (\ref{tab:mse.kernel}) here below report the 
finite sample performance of stopped boosted smoother by the GCV criterion. 
Each entry in the table reports the median number of iterations and the median 
mean square error over 100 simulations. As expected,  larger 
smoothing parameter of the initial smoother require more iterations
of the boosting algorithm to reach its optimum.   Interestingly, the selected
smoother starting with a very smooth smoother, has slightly smaller 
mean squared error.    The quantify the benefits of the iterative bias correction 
scheme, the last column of the tables gives the mean squared error of the
original smoother with smoothing parameters selected using GCV.
In all cases, the iterative bias correction has smaller mean squared error
than the "one-step" smoother,  with improvements ranging from 15\% to 30\%.

Table (\ref{tab:mse}) presents the results for smoothing splines.
\begin{table}[H]
\centering
\begin{tabular}[b]{cccccccc}\hline\hline
\multicolumn{8}{c}{Function $m_1$}\\\hline
error&
$\hat k_1$1&
$S_{\hat k_{\mathrm{GCV}}}(\lambda_1)$&
$\hat k_2$&
$S_{\hat k_{\mathrm{GCV}}}(\lambda_2)$&
$\hat k_3$&
$S_{\hat k_{\mathrm{GCV}}}(\lambda_3)$&
$S(\hat \lambda_\mathrm{GCV})$\\
\hline
Gaussian&4077&0.0273&65&0.0282&2&0.0293&0.0379\\
student &4115&0.0273&70&0.0286&2&0.0296&0.0352\\
\hline
\multicolumn{8}{c}{Function $m_2$}\\
\hline
Gaussian&1219&0.0798&21&0.0845&1&0.0837&0.0829\\
student &1307&0.0887&22&0.0944&1&0.0932&0.0937\\
\hline
\multicolumn{8}{c}{Function $m_3$}\\
\hline
Gaussian&135&0.0014&3&0.0014&1&0.0016&0.0016\\
student &147&0.0016&3&0.0016&1&0.0018&0.0019\\
\end{tabular}
\caption{Median over 100 simulations of the number of iterations and the
MSE for smoothing splines smoother, $n=50$ data points. 
$S_(\hat \lambda_\mathrm{GCV})$ denotes the traditional smoothing splines estimate 
with $\lambda$ chosen with GCV.}
\label{tab:mse}
\end{table}

Table (\ref{tab:mse.kernel}) presents the results for kernel smoothers
with a Gaussian kernel.

\begin{table}[H]
\centering
\begin{tabular}[b]{cccccccc}\hline\hline
\multicolumn{8}{c}{Function $m_1$}\\\hline
error&
$\hat k_1$1&
$S_{\hat k_{\mathrm{GCV}}}(h_1)$&
$\hat k_2$&
$S_{\hat k_E{\mathrm{GCV}}}(h_2)$&
$\hat k_3$&
$S_{\hat k_{\mathrm{GCV}}}(h_3)$&
$S(\hat h_\mathrm{AICc})$\\\hline
Gaussian&385&0.0231&27&0.0254&4&0.0368&0.04857\\
student &360&0.0221&25&0.0262&4&0.0353&0.05199\\\hline
\multicolumn{8}{c}{Function $m_2$}\\\hline
Gaussian&330&0.0477&128&0.0581&14&0.0782&0.1175\\
student &1621&0.0563&160&0.0660&16&0.0754&0.1184\\\hline
\multicolumn{8}{c}{Function $m_3$}\\\hline
Gaussian&30&0.0017&7&0.0016&2&0.0014&0.00178\\
student &29&0.0017&8&0.0016&2&0.0016&0.0018\\
\end{tabular}
\caption{Median over 100 simulations of the number of iterations and
the MSE for Gaussian kernel smoother, $n=50$ data points. 
$S_(\hat h_\mathrm{AICc})$ denotes the bandwidth chosen by the modified AIC 
criteria.}
\label{tab:mse.kernel}
\end{table}

The simulation results reported in the above tables show that
the iterative bias reduction scheme works well in practice, even 
for moderate sample sizes.  While starting with a very smooth
pilot requires more iterations, the mean squared 
error of the resulting smoother is somewhat smaller compared
to a more noisy initial smoother.  Figures 
\ref{fig:initialsmoother} and \ref{fig:finalsmoother}
also support this claim.

\section{Discussion}
In this paper, we make the connection between iterative bias correction
and the $L_2$ boosting algorithm, thereby providing a new interpretation 
for the latter.   A link between bias reduction and boosting was suggested by  
 \cite{ridgeway2000} in his discussion of the seminal paper
 \cite{friedman++2000}, and explored in \citet{marzio+2004,marzio+2007}
 for the special case of kernel smoothers.
 In this paper, we show that this interpretation holds for general linear 
 smoothers.   
 
 It was surprising to us that not all smoothers were suitable to be used
 for boosting.  We show that many weak learners, such as the
 $k$-nearest neighbor smoother and some kernel smoothers, are not stable
 under $L_2$ boosting.  Our results extend and complement the
 recent results of \citet{marzio+2007}.  
 
 Iterating the boosting algorithm until convergence is not desirable.
 Better smoothers result if one stops the iterative scheme.  We have
 explored, via simulations, various data driven stopping rules and have
 found that for the linear smoothers, the Generalized Cross Validation
 criteria works very well, even for moderate sample sizes of $50$.
 In our simulations show that optimally correcting the bias (by
 stopping the $L_2$ boosting algorithm after a suitable number of iterations)  
 produced better smoothers than the one with the best data-dependent 
 smoothing parameter.
 
Finally,  the iterative bias correction scheme can be readily extended 
to multivariate covariates $X$, as in \citet{buhlmann2006}.

\appendix
\section{Appendix}
{\bf Proof of Theorem \ref{theorem:iterative}}  To show \ref{eq:fk}, let
$\Sigma = I + (I-S) + \dots + (I-S)^{k-1}$.   The conclusion follows
from a telescoping sum argument applied to
\[
S\Sigma = \Sigma - (I-S) \Sigma = I - (I-S)^k.
\]

{\bf Proof of Theorem \ref{theorem:converge}}
\begin{eqnarray*}
\|\hat b_k\|^2 &=& \|-(I-S)^{k-1} SY\|^2\\
&=&  \|(I-S)(I-S)^{k-2} SY\|^2  \leq \|(I-S)\|^2 \|\hat b_{k-1}\|^2\\
& \leq & \|\hat b_{k-1}\|^2,
\end{eqnarray*}
where the last inequality follows from the assumptions on the spectrum of $I-S$.
Similarly,  one shows that

\begin{eqnarray*}
\|R_k\|^2 &=& \| (I-S)^kY\|^2 \leq \|I-S\|^2 \|R_{k-1}\|^2  < \|R_{k-1}\|^2.
\end{eqnarray*}

{\bf Proof of Theorem \ref{knn}}
To simplify the exposition, let us assume that the $X_i$'s are ordered. 
Let us consider the $K$-nn smoother the matrix $S$ is of general term 
\begin{eqnarray*}
S_{ij} &=& \frac{1}{K} \quad \hbox{if} \quad X_j \in \text{$K$-nn}(X_i).
\end{eqnarray*}
In order to bound the singular values of $(I-S)$, consider the 
eigen values of $(I-S)(I-S)'$ which are the square of the singular values of 
$I-S$.
Since $A=(I-S)(I-S)'$ is symmetric,  we have for any vector $x$ that 
\begin{eqnarray}
\label{bornevp}
\lambda_n \leq \frac{x' A x}{x' x} \leq \lambda_1.
\end{eqnarray} 
Let us find a vector $x$ such that $x' A x > x'x$. First notice that 
\begin{eqnarray*}
A_{ii} &=& 1 - \frac{1}{K}.
\end{eqnarray*}
Next, consider the vector $x$ of $\R^n$ that is zero every where except 
at position $(i-l_1)$ (respectively $i$ and $i+l_2$) where its value is 
-1 (respectively 2 and -1). For this  choice, we expand $x' A x$ to get
\begin{eqnarray*}
x' A x &=& A_{i-l_1,i-l_1}+4 A_{i,i} +A_{i+l_2,i+l_2} 
-4 A_{i-l_1,i} -4 A_{i,i+l_2} +2 A_{i+l_2,i-l_1}\\
&=& 6 - \frac{6}{K}-4 A_{i-l_1,i} -4 A_{i,i+l_2} +2 A_{i+l_2,i-l_1}.
\end{eqnarray*}
To show that this last quantity is larger than $x^t x = 6$, we need to suitably
bound the off-diagonal elements of $A = I - S- S' +SS'$.
To bound $A_{ij}$, where $j=i+l$ and $l<K$, we need to consider three cases:
\begin{enumerate}
\item If $X_i$ belongs to the $K$-nn of $X_j$ and vice versa, then $S_{ij} = S_{ji}'=1/K$.
This does not mean that all the $K$-nn neighbor of $X_i$ are the same as those 
of $X_j$, but if it is the case, then $(SS')_{ij} \leq K/K^2$ and otherwise in the 
pessimistic case, we bound $(SS')_{ij} \geq (l+1)/K^2$.  It therefore follows that
\begin{eqnarray*}
(l+1)/K^2 - \frac{2}{K} \leq &A_{i,i+l} &\leq \frac{K}{K^2}-\frac{2}{K} = -\frac{1}{K}.
\end{eqnarray*}
\item If $X_i$ belongs to the $K$-nn of $X_j$ $S_{ij} =1/K$ but $X_j$ does not 
belong to the $K$-nn of $X_i$  then $S_{ji}'=0$. There is at a maximum of
$K-1$ points that are in the $K$-nn of $X_i$ and in the $K$-nn of $X_j$ so
$(SS')_{ij} \leq (K-1)/K^2$. In the pessimistic case, there is only one 
point, which leads to the bound 
\begin{eqnarray*}
\frac{1}{K^2} -\frac{1}{K} \leq &A_{i,i+l} &
\leq \frac{K-1}{K^2}-\frac{1}{K} \leq - \frac{1}{K^2}.
\end{eqnarray*}
\item If $X_i$ does not belong to the $K$-nn of $X_j$ $S_{ij}=0$ and $X_j$ does not 
belong to the $K$-nn of $X_i$  then $S_{ji}'=0$. However there are potentially 
as many as $l-1$ points that are in the $K$-nn of $X_i$ and in the $K$-nn of $X_j$, 
so that $(SS')_{ij} \leq (l-1)/K^2$. In that case 
\begin{eqnarray*}
0 \leq A_{ij} &\leq& \frac{l-1}{K^2}\leq \frac{K-2}{K^2}.
\end{eqnarray*}
\end{enumerate}
With these bounds for the off-diagonal terms, we are able 
to major $x'Ax$. 

Before continuing, we need to discuss the relative
position of the points $X_{i-l_1}, X_i$ and $X_{i+l_2}$. We chose them such that
\[ 
X_{i-l_1} \in \text{$K$-nn}(X_i) \quad \mbox{and} \quad X_i \in \text{$K$-nn}(X_{i-l_1}).
\]
For this choice, we calculate 
\begin{eqnarray*}
\frac{l_1+1-2K}{K^2} \leq & A_{i-l_1,i} & \leq -\frac{1}{K}\\
\frac{l_2+1-2K}{K^2} \leq & A_{i,i+l_2} & \leq -\frac{1}{K},\\
\end{eqnarray*}
so that
\begin{eqnarray*}
6 - \frac{6}{K} + \frac{8}{K}+ 2 A_{i+l,i-l} \leq x' A x \leq
 6 + \frac{2}{K} + 2 A_{i+l,i-l}.
\end{eqnarray*}
The latter shows that $x'Ax > x'x$ whenever 
\begin{eqnarray*}
A_{i+l_2,i-l_1} >  -  \frac{1}{K},
\end{eqnarray*}
which is always true if the condition 
\[
X_{i-l_1} \not \in \text{$K$-nn}(X_{i+l_2}) \quad \mbox{or} \quad
X_{i+l_2} \not \in \text{$K$-nn}(X_{i-l_1})
\]
is satisfied because in such case, we have
\begin{eqnarray*}
-\frac{1}{K}  < &A_{i-l_1,i+l_2} & \leq  \frac{1}{K^2}.
\end{eqnarray*}

{\bf Proof of Theorem \ref{kernel}} Let $X_1,\ldots,X_n$ is an i.i.d.
sample from a density $f$ that is bounded away from zero on a compact
set strictly included in the support of $f$. Consider 
without loss of generality that $f(x) \geq c > 0$ for all $|x| < b$.

We are interested in the sign of the quadratic form $u^tAu$ where 
the individual entries $A_{ij}$ of matrix $A$ are equal to
\begin{eqnarray*}
A_{ij} &=&  \frac{K_h(X_i-X_j)}{\sqrt{\sum_l K_h(X_i-X_l)}\sqrt{\sum_lK_h(X_j-X_l)}}.
\end{eqnarray*}
Recall the definition of the scaled kernel $K_h(\cdot) = K(\cdot/h)/h$.
If $v$ is the vector of coordinate 
$v_i=u_i/\sqrt{\sum_l K_h(X_i-X_l)}$ then we have $u^t Au=v^t \mathbb{K} v$, where $\mathbb{K}$ is the matrix with individual entries $K_h(X_i-X_j)$. Thus
any conclusion on the quadratic form
$v^t\mathbb{K}v$ carry on to the quadratic form $u^tAu$. 

\medskip

To show the existence of a negative eigenvalue for ${\mathbb K}$,
we seek to construct a vector $U=(U_1(X_1),\ldots,U_n(X_n))$ 
for which we can show that the quadratic form
\[
U^t {\mathbb K} U = \sum_{j=1}^n \sum_{k=1}^n U_j(X_j) U_k(X_k) K_h(X_j-X_k)
\]
converges in probability to a negative quantity as the
sample size grows to infinity.   We show the latter by
evaluating the expectation of the quadratic form and 
applying the weak law of large number.

Let $\varphi(x)$ be a real function in $L_2$, define its Fourier 
transform 
\begin{eqnarray*}
\hat \varphi(t) &=& \int e^{-2i\pi t x} \varphi(x) dx
\end{eqnarray*}
and its Fourier inverse by
\begin{eqnarray*}
\hat \varphi_{inv}(t) &=& \int e^{2i\pi t x} \varphi(x) dx.
\end{eqnarray*}
For kernels $K(\cdot)$ that are real symmetric  probability densities, 
we have
\begin{eqnarray*}
\hat K(t) &=& \hat K_{inv}(t).
\end{eqnarray*}

From Bochner's theorem, we know that if the kernel $K(\cdot)$ is not positive
definite, then there exists a bounded symmetric set $A$ of positive 
Lebesgue measure (denoted by $|A|$), such that 
\begin{equation} 
\label{eq:not.pd}
\hat K(t) < 0 \quad \forall t \in A.
\end{equation}
Let $\widehat \varphi(t)\in L_2$ be a real symmetric function
supported on $A$, bounded by $B$ (i.e. $|\widehat \varphi(t)|
\leq B$). Obviously, its inverse Fourier transform 
\[
\varphi(x) = \int_{-\infty}^\infty e^{-2\pi i x t} \widehat \varphi(t) 
dt
\]
is integrable and by virtue of Parceval's identity
\begin{eqnarray*}
\|\varphi\|^2 = \|\widehat \varphi\|^2 \leq B^2 |A| < \infty.
\end{eqnarray*}
Using the following version of Parceval's identity \citep[see][p.620]{feller1966}
\begin{equation*} \label{eq:parceval}
\int_{-\infty}^\infty \int_{-\infty}^\infty \varphi(x) \varphi(y) K(x-y) dx dy 
= \int_{-\infty}^\infty |\widehat \varphi(t)|^2 \hat K(t) dt,
\end{equation*}
which when combined with equation (\ref{eq:not.pd}), leads us to conclude that
\begin{eqnarray*}
\int_{-\infty}^\infty \int_{-\infty}^\infty \varphi(x) \varphi(y) K(x-y) dx dy < 0.
\end{eqnarray*}

Consider 
the following vector
\[
U = \frac{1}{nh} \left [ 
\begin{array}{c}  
\frac{\varphi(X_1/h)}{f(X_1)} {\mathbb I}(|X_1| < b)\\
\frac{\varphi(X_2/h)}{f(X_2)} {\mathbb I}(|X_2| < b)\\
\vdots\\
\frac{\varphi(X_n/h)}{f(X_n)} {\mathbb I}(|X_n| < b)\\
\end{array}
\right ].
\] 
With this choice, the expected value of the quadratic form is 
\begin{eqnarray*}
\E[Q] & = & \E \left [ \sum_{j,k=1}^n U_j(X_j) U_k(X_k) K_h(X_j-X_k)\right ]\\
&=& \frac{1}{n} \int_{-b}^b \frac{1}{f(s) h^2} \varphi(s/h)^2 K_h(0) ds \\
&&\quad +
\frac{n^2-n}{n^2} \int_{-b}^b \int_{-b}^b \frac{1}{h^2}
\varphi(s/h)\varphi(t/h) K_h(s-t) ds dt\\
&=& I_1 + I_2.
\end{eqnarray*}
We bound the first integral
\begin{eqnarray*}
I_1 & = & \frac{K_h(0)}{nh^2} \int_{-b}^b \frac{\varphi(s/h)^2}{f(s)} ds\\
&\leq& \frac{K_h(0)}{nch} \int_{-b/h}^{b/h} \varphi(u)^2 du\\
&\leq& \frac{B^2 |A| K(0)}{c h^2} n^{-1}.
\end{eqnarray*}
Observe that for any fixed value $h$, the latter can be made arbitrarily small
by choosing $n$ large enough.   We evaluate the second integral by noting 
that
\begin{eqnarray}
I_2 &=& \left ( 1-\frac{1}{n} \right ) h^{-2} \int_{-b}^b \int_{-b}^b \varphi(s/h)
\varphi(t/h) K_h (s-t) ds dt \nonumber \\
&=& \left ( 1-\frac{1}{n} \right ) h^{-2} \int_{-b}^b \int_{-b}^b \varphi(s/h)
\varphi(t/h) \frac{1}{h} K\left ( \frac{s}{h} - \frac{t}{h} \right ) ds dt \nonumber \\
&=& \left ( 1-\frac{1}{n} \right ) h^{-1} \int_{-b/h}^{b/h} \int_{-b/h}^{b/h}
\varphi(u) \varphi(v) K(u-v) du dv. \label{eq:want.negative}
\end{eqnarray}

By virtue of the dominated convergence theorem, the value of the 
last integral converges to $\int_{-\infty}^\infty |\widehat \varphi(t)|^2 
\hat K(t) dt < 0$ as $h$ goes to zero.  Thus for $h$ small enough, 
(\ref{eq:want.negative}) is less than zero,  and it follows that 
we can make ${\mathbb E}[Q] < 0$ by taking $n \geq n_0$, for some large 
$n_0$.   Finally, convergence in probability of the quadratic form to its 
expectation is guaranteed by the weak law of large numbers for $U$ statistics
(see \citet{grams+1973} for example).   The conclusion of the theorem follows.

{\bf Proof of Proposition \ref{unif}}
We are interested in the sign of the quadratic form $u^t\mathbb{K}u$ (see proof of Theorem \ref{kernel}).  Recall that if $\mathbb{K}$ is semidefinite 
then all its principal minor \citep[see][p.398]{horn+1985} 
are nonnegative.   In particular, we can show that $A$ is non-positive definite
by producing a $3 \times 3$ principal minor with negative determinant.
To this end, take the principal minor $\mathbb{K}[3]$ obtained by taking the rows 
and columns $(i_1,i_2,i_3)$.  Without loss of generality, let us assume that
$X_{i_1}<X_{i_2}<X_{i_3}$.   The determinant of $\mathbb{K}[3]$ is
\begin{eqnarray*}
det(\mathbb{K} [3])
&=&\quad K_h(0)\left[K_h(0)^2 - K_h(X_{i_3}-X_{i_2})^2\right]\\
&& \!\!\quad - K_h(X_{i_2}-X_{i_1})\\
&& \!\!\quad \quad \times \left[K_h(0)K_h(X_{i_2}-X_{i_1})-K_h(X_{i_3}-X_{i_2})K_h(X_{i_3}-X_{i_1})\right]\\
&& \!\!\quad + K_h(X_{i_3}-X_{i_1})\\
&& \!\!\quad \quad \times
\left[K_h(X_{i_2}-X_{i_1})K_h(X_{i_3}-X_{i_2})-K_h(0)K_h(X_{i_3}-X_{i_1})\right].
\end{eqnarray*}
Let us evaluate this quantity for the uniform and Epanechnikov kernels.\\
\\
\noindent
{\bf Uniform kernel.}  Let $h$ be larger than the minimum 
distance between three consecutive points, and chose the index 
$i_1,i_2,i_3$ such that 
\[
X_{i_2}-X_{i_1} < h, \quad X_{i_3}-X_{i_2} < h, \quad \mbox{and} \quad
X_{i_3} - X_{i_1} > h.
\]
With this choice, we readily calculate
\begin{eqnarray*}
det(\mathbb{K} [3])&=&0-K_h(0)\left[K_h(0)^2-0\right]-0<0.
\end{eqnarray*}
Since a principal minor of $\mathbb{K}$ is negative, we conclude that
$\mathbb{K}$ and $A$ are not semidefinite positive.\\
\\
\noindent
{\bf Epanechnikov kernel.} For $i_1,i_2,i_3$ fixed, 
denote by $x=X_{i_2}-X_{i_1}$ and by 
$y=X_{i_3}-X_{i_2}$, and assume that 
$h> \min(x,y)$.  The determinant $det(\mathbb{K} [3])$ is 
a bivariate function of $x$ and $y$ (as $X_{i_3}-X_{i_1} = x+y$).
Numerical evaluations of that function show that as soon 
as we have the range of the three points less than the bandwidth, 
the determinant of $\mathbb{K} [3]$ is negative.
\begin{figure}[H]
\begin{center}
\includegraphics{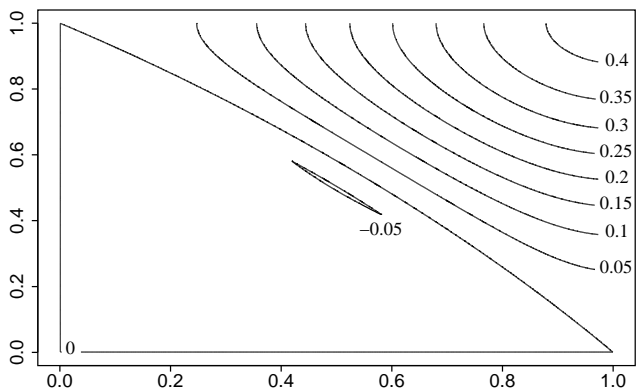}
\caption{Contour of $det(\mathbb{K} [3])$ as a function of $(x,y)$.}
\end{center}
\end{figure}
Thus a principal minor of $\mathbb{K}$ is negative, and as a result,  
$\mathbb{K}$ and $A$ are not semidefinite positive.

\bibliographystyle{abbrvnat} 
\bibliography{./biblio}
\end{document}